\documentclass[12pt]{article}
\makeatletter
\newcommand{\singlespacing}{\let\CS=\@currsize\renewcommand{\baselinestretch}{1}\tiny\CS}
\usepackage{epsfig}
\usepackage{amsbsy}
\usepackage{amssymb,color}
\begin{document}
\baselineskip=24pt
\parskip = 10pt
\def \qed {\hfill \vrule height7pt width 5pt depth 0pt}
\newcommand{\ve}[1]{\mbox{\boldmath$#1$}}
\newcommand{\IR}{\mbox{$I\!\!R$}}
\newcommand{\1}{\Rightarrow}
\newcommand{\bs}{\baselineskip}
\newcommand{\esp}{\end{sloppypar}}
\newcommand{\be}{\begin{equation}}
\newcommand{\ee}{\end{equation}}
\newcommand{\beanno}{\begin{eqnarray*}}
\newcommand{\inp}[2]{\left( {#1} ,\,{#2} \right)}
\newcommand{\eeanno}{\end{eqnarray*}}
\newcommand{\bea}{\begin{eqnarray}}
\newcommand{\eea}{\end{eqnarray}}
\newcommand{\ba}{\begin{array}}
\newcommand{\ea}{\end{array}}
\newcommand{\nno}{\nonumber}
\newcommand{\dou}{\partial}
\newcommand{\bc}{\begin{center}}
\newcommand{\ec}{\end{center}}
\newcommand{\2}{\subseteq}
\newcommand{\cl}{\centerline}
\newcommand{\ds}{\displaystyle}
\def\refhg{\hangindent=20pt\hangafter=1}
\def\refmark{\par\vskip 2.50mm\noindent\refhg}

\title{\sc  On Bivariate Discrete Weibull Distribution}

\author{Debasis Kundu\footnote{Department of Mathematics and Statistics, Indian Institute of Technology Kanpur, Kanpur,
Pin 208016, \ \ \ \ \ \ \ \ India.  \ \ e-mail: kundu@iitk.ac.in.} \& Vahid Nekoukhou\footnote{Department of Statistics, Khansar Faculty of Mathematics and Computer Science, Khansar, Iran.}}

\date{}
\maketitle

\begin{abstract}
Recently, Lee and Cha (2015, `On two generalized classes of discrete bivariate distributions', {\it American Statistician}, 221 - 230)
proposed two general classes of discrete bivariate distributions.  They have discussed some general properties and some specific cases
of their proposed distributions.  In this paper we have considered one model, namely bivariate discrete Weibull distribution,
which has not been considered in the literature yet.  The proposed bivariate discrete Weibull distribution is a discrete analogue of
the Marshall-Olkin bivariate Weibull distribution.  We study various properties of the proposed distribution and discuss its interesting 
physical interpretations.  The proposed model has four parameters, and because of that it is a very flexible
distribution.  The maximum likelihood estimators of the parameters cannot be obtained in closed forms, and we have proposed 
a very efficient nested EM algorithm which works quite well for discrete data.  We have also proposed augmented Gibbs sampling
procedure to compute Bayes estimates of the unknown parameters based on a very flexible set of priors.
Two data sets have been analyzed to show how the proposed model and the method work in practice.  We will see that the performances are
quite satisfactory.  Finally, we conclude the paper.

\end{abstract}





\newpage

\section{\sc Introduction}

Analyzing discrete bivariate data is quite common in practice.  Discrete bivariate data arise quite naturally in many real
life situations and are often highly correlated.  For example, the number of goals scored by two
competing teams or the number of insurance claims for two different causes is an example of typical discrete bivariate data.
Several bivariate discrete distributions are available in the literature.  Encyclopedic surveys of different discrete bivariate
distributions can be found in Kocherlakota and Kocherlakota \cite{KK:1992} and Johnson et al. \cite{JKB:1997}, see also Ong and Ng \cite{ON:2013}, Nekoukhou and Kundu \cite{NK:2017}, Kundu and Nekoukhou \cite{KN:2018} and the references cited therein.

Recently, Lee and Cha \cite{LC:2015} proposed two fairly general classes of discrete bivariate distributions based on the minimization
and maximization methods.  They discussed some specific cases namely bivariate Poisson, bivariate geometric, bivariate negative
binomial and bivariate binomial distributions.  Although, the method proposed by Lee and Cha \cite{LC:2015} is a very powerful method, the
joint probability mass function (PMF) may not be always a convenient form.  Moreover, the bivariate distributions proposed by Lee and Cha
\cite{LC:2015} may not have the same corresponding univariate marginals.  For example, the bivariate Poisson and bivariate geometric
distributions do not have univariate Poisson and univariate geometric marginals, respectively.  This may not be very desirable.
Moreover, Lee and Cha \cite{LC:2015} also did not discuss any inferential issues of the unknown parameters.

Nakagawa and Osaki \cite{NO:1975} introduced the discrete Weibull (DW) distribution, which can be considered as a discrete
analogue of the absolutely continuous Weibull distribution.  The hazard function of the DW distribution can be increasing,
decreasing or constant depending on its shape parameter.  The geometric distribution can be obtained as a special case. The DW
distribution has been used quite successfully in different areas, see for example in population dynamics
(e.g. Wein and Wu \cite{WW:2001}), stress-strength reliability (e.g. Roy \cite{Roy:2002}), evaluation of reliability of complex
systems (e.g. Roy \cite{Roy:2002}), wafer probe operation in semiconductor manufacturing (e.g. Wang \cite{Wang:2009}), minimal
availability variation design of repairable systems (e.g., Wang et al. \cite{WYYZ:2010}) and microbial counts in water
(e.g. Englehardt and Li \cite{EL:2011}).

The main aim of the present paper is to consider the bivariate discrete Weibull (BDW) distribution which can be obtained from
three independent DW distributions by using the minimization method.  It can be considered as a natural discrete analogue of the
Marshall-Olkin bivariate Weibull (MOBW) distribution, see for example Marshall and Olkin \cite{MO:1967} or Kundu and Dey \cite{KD:2009}
for detailed description of the MOBW distribution.  The BDW distribution is a very flexible bivariate discrete 
distribution, and its joint PMF depending on the parameter values can take various shapes. The generation from a BDW
distribution is straight forward, and hence the simulation experiments can be performed quite conveniently.
It has also some interesting physical interpretations. In addition, its marginals are DW distributions. Hence, a new bivariate
distribution is introduced whose marginals are able to analyze the monotone hazard rates in the discrete case. In addition, a
new three-parameter bivariate geometric distribution can be obtained as a special case.

We have provided several properties of the proposed BDW distribution.  It has some interesting physical interpretations in terms of
the discrete shock model and latent failure time competing risks model. The BDW distribution has four unknown parameters.  The maximum likelihood estimators (MLEs) cannot be obtained in explicit forms.  The MLEs can be obtained after solving four non-linear equations.  The standard algorithms like
Newton-Raphson method may be used to compute the MLEs.  Since it involves solving four non-linear equations simultaneously, it has the
standard problems of choosing the efficient initial guesses and the convergence of the algorithm to a local minimum rather than a
global minimum.  To avoid that problems we treat this problem as a missing value problem, and provided a very efficient expected
maximization (EM) algorithm to compute the MLEs.  We further consider the Bayesian inference of the unknown parameters.  It is assumed
that the scale parameters have a very flexible Dirichlet-gamma prior and the shape parameter has a prior with a log-concave probability
density function (PDF).  The Bayes estimators of the unknown parameters cannot be obtained in explicit forms in general and we have
used Gibbs sampling technique to compute the Bayes estimates and the associated highest posterior density credible intervals.  Two real 
data sets; (i) Italian football score data and (ii) Nasal drainage severity score data, have been analyzed for illustrative purposes
 mainly to see how the proposed 
model and the methods perform in practice.  The performances are quite satisfactory.

The rest of the paper is organized as follows.  In Section 2, we have provided the preliminaries and the priors.  Different basic properties
are discussed in Section 3.  In Sections 4 and 5, we have considered the classical and Bayesian inference, respectively.  The
analysis of two real data sets have been presented in Section 6, and finally we conclude the paper in Section 7.

\section{\sc Preliminaries and Prior Assumptions}

\subsection{\sc The Weibull and DW Distributions}
Weibull \cite{Weibull:1951} introduced an absolutely continuous distribution that plays a key role in reliability studies. The
cumulative distribution function (CDF) and the PDF of the Weibull distribution with the shape parameter
$\alpha>0$ and the scale parameter $\lambda>0$ are
\bea
F_{WE}(x; \alpha, \lambda) & = & 1-e^{-\lambda x^\alpha}, \quad x>0, \ \ \ \ \hbox{and}  \nonumber  \\
f_{WE}(x; \alpha, \lambda) & = & \alpha \lambda x^{\alpha-1}e^{-\lambda x^\alpha}, \quad x>0,
\eea
respectively.
From now on WE$(\alpha,\lambda)$ is used to represent a Weibull distribution with the shape parameter $\alpha$ and the scale
parameter $\lambda$. The Weibull distribution is a generalization of the exponential distribution and hence the exponential
distribution is obtained as a special case (when $\alpha=1)$.  The PDF and hazard rate function of the Weibull distribution
can take various shapes.  The PDF can be a decreasing or an unimodal function and the hazard rate function can be an increasing (when $\alpha>1$), decreasing (when $\alpha<1$) or a constant function (when $\alpha=1$).  For a detailed discussions on Weibull
distribution one is referred to the book length treatment by Johnson et al. \cite{JKB:1995}.

As mentioned before, Nakagawa and Osaki \cite{NO:1975} introduced the discrete Weibull distribution, which can be considered as a discrete analogue of the absolutely continuous Weibull distribution. The PMF of a DW distribution with parameters $\alpha>0$ and $0<p<1$, is given by
\begin{eqnarray}
f_{DW}(y;\alpha,p)=p^{y^\alpha}-p^{(y+1)^\alpha}, \quad y\in\mathbb{N}_0=\{0,1,2,...\}.
\end{eqnarray}
DW$(\alpha, p)$ is used to represent a DW distribution in the sequel. The survival function (SF) of a DW$(\alpha,p)$ is also given by
\begin{eqnarray}
S_{DW}(y;\alpha,p)=P(Y\geq y)=p^{ [y]^\alpha}.
\end{eqnarray}
Here, $[y]$ denotes the largest integer less than or equal to $y$.

\noindent {\sc Proposition 1:} Let $X_1,X_2,...,X_n$ be a random sample from a DW$(\alpha, p)$ distribution. Then, $\min\{X_1,X_2,...,X_n\}\sim$ DW$(\alpha,p^n)$.

\noindent {\sc Proof.} The proof is straight forward and the details are avoided. \qed

The following representation of a DW random variable becomes very useful. If $Y \sim$ W$(\alpha,\lambda)$, then for
$\ds p=e^{-\lambda}$,
\be
X = [Y] \sim \hbox{DW}(\alpha, p).    \label{repre}
\ee
Using (\ref{repre}), the generation of a random sample from a DW$(\alpha,p)$ becomes very simple. More precisely, first we can generate
a random sample $X$ from a WE$(\alpha,\lambda)$ distribution, and then by considering $Y = [X]$, we can obtain a generated sample from
DW$(\alpha,p)$.

\subsection{\sc Marshall-Olkin Bivariate Weibull Distribution}

Marshall and Olkin \cite{MO:1967} proposed the MOBW distribution as follows.  Suppose $U_0, U_1$ and $U_2$ are three independent random variables,
such that
\be
U_0 \sim \hbox{WE}(\alpha, \lambda_0), \ \ \ U_1 \sim \hbox{WE}(\alpha, \lambda_1) \ \ \ \hbox{and} \ \ \ U_2 \sim \hbox{WE}(\alpha, \lambda_2).
\ee
Here `$\sim$' means follows in distribution.  Then the random variables $(Y_1, Y_2)$, where
$$
Y_1 = \min\{U_0, U_1\} \ \ \ \hbox{and} \ \ \ Y_2=\min\{U_0, U_2\},
$$
is known to have MOBW distribution with parameters $\alpha, \lambda_0, \lambda_1$ and $\lambda_2$.  The joint survival function of $Y_1$ and
$Y_2$ can be written as
\be
S_{Y_1, Y_2}(y_1, y_2) = P(Y_1 > y_1, Y_2 > y_2) = e^{-\lambda_1 y_1^{\alpha} - \lambda_2 y_2^{\alpha} - \lambda_0 [\max \{y_1, y_2\}]^{\alpha}},
\ee
for $y_1 > 0$ and $y_2 > 0$.  The joint PDF can be written as
\bea
f_{Y_1, Y_2}(y_1, y_2) = \left \{
\begin{array}{lll}
f_{WE}(y_1; \alpha, \lambda_1) f_{WE}(y_2; \alpha, \lambda_0+\lambda_2) & \hbox{if} & y_1 < y_2 \cr
f_{WE}(y_1; \alpha, \lambda_0+\lambda_1) f_{WE}(y_2; \alpha, \lambda_2) & \hbox{if} & y_1 > y_2 \cr
\frac{\lambda_0}{\lambda_0+\lambda_1+\lambda_2} f_{WE}(y; \alpha, \lambda_0+\lambda_1+\lambda_2) & \hbox{if} & y_1 = y_2 = y,
\end{array}
\right .
\eea
see Kundu and Dey \cite{KD:2009} for details.  From now on it will be denoted by MOBW$(\alpha, \lambda_0, \lambda_1, \lambda_2)$.

\subsection{\sc Prior Assumptions}

Kundu and Gupta \cite{KG:2013} provided the Bayesian analysis of the MOBW distribution based on the following prior assumptions.
When the common shape parameter $\alpha$ is known, it is assumed that the joint prior of $\lambda_0$, $\lambda_1$ and $\lambda_2$ is
\bea
\pi_1(\lambda_0, \lambda_1, \lambda_2|a,b,a_0,a_1,a_2) & = & \frac{\Gamma(a_0+a_1+a_2)}{\Gamma(a)} (b \lambda) ^{a-a_0-a_1-a_2}
\frac{b^{a_0}}{\Gamma(a_0)} \lambda_0^{a_0-1} e^{-b \lambda_0}  \nonumber \\
& & \times \frac{b^{a_1}}{\Gamma(a_1)} \lambda_1^{a_1-1} e^{-b \lambda_1} \times \frac{b^{a_2}}{\Gamma(a_2)} \lambda_2^{a_2-1} e^{-b \lambda_2},
\label{prior}
\eea
for $0 < \lambda_0, \lambda_1, \lambda_2$.  Here $0 < a, b, a_0, a_1, a_2 < \infty$ are all hyper-parameters and $\lambda = \lambda_0+\lambda_1
+\lambda_2$.  The prior (\ref{prior}) is known as the Dirichlet-Gamma prior, and from now on it will be denoted by
DG$(a,b,a_0,a_1,a_2)$.  It may be mentioned that Pena and Gupta \cite{PG:1990} first considered this prior in case of 
the Marshall-Olkin bivariate exponential distribution and discussed its different properties.  It has been shown that all the
parameters are identifiable and estimable also.
It is a very flexible prior, and depending on the values of the hyper-parameters $\lambda_i$ and $\lambda_j$ for
$i \ne j$, can be independent, positively or negatively correlated.  For known $\alpha$, it is a conjugate prior.  When the shape parameter $\alpha$ is not known, Kundu and Gupta \cite{KG:2013} did not assume any specific form of the
prior on $\alpha$.  It is simply assumed that the prior of $\alpha$ has a non-negative support on $(0, \infty)$, and the PDF
of the prior of $\alpha$, say $\pi_2(\alpha)$, is log-concave.  Moreover, $\pi_1(\cdot)$ and $\pi_2(\cdot)$ are independently
distributed.  In this paper we have also assumed the same set of priors, and the details will be explained later.

\section{\sc The BDW Distribution and its Properties}

\subsection{\sc Definition and Interpretations}

\noindent {\sc Definition:}  Suppose $U_1\sim$ DW$(\alpha,p_1)$,  $U_2 \sim$ DW$(\alpha,p_2)$ and $U_0 \sim$ DW$(\alpha,p_0)$ and
they are independently distributed. If $X_1=\min\{U_1,U_0\}$ and $X_2=\min\{U_2,U_0\}$, then we say that the bivariate vector
$(X_1,X_2)$ has a BDW distribution with parameters $\alpha$, $p_0$, $p_1$ and $p_2$.  From now on we denote this bivariate discrete distribution by
BDW$(\alpha, p_0, p_1, p_2)$.

\noindent If $(X_1,X_2)\sim$ BDW$(\alpha, p_0, p_1, p_2)$, then the joint SF of $(X_1,X_2)$ for $x_1 \in \mathbb{N}_0$,
$x_2 \in \mathbb{N}_0$ and for $z=\max\{x_1,x_2\}$ is
\begin{eqnarray}\nonumber
S_{X_1,X_2}(x_1,x_2)&=&P(X_1\geq x_1, X_2\geq x_2) =  p_1^{{x_1}^{\alpha}}p_2^{{x_2}^{\alpha}}p_0^{z^{\alpha}} \\
&=&S_{DW}(x_1; \alpha, p_1)S_{DW}(x_2; \alpha, p_2)S_{DW}(z; \alpha, p_0). \nonumber
\label{r6}
\end{eqnarray}
The joint SF of $(X_1,X_2)$ can also be written as
\begin{eqnarray}
S_{X_1,X_2}(x_1,x_2)= \left\{%
\begin{array}{lll}
S_{DW}(x_1; \alpha, p_1)S_{DW}(x_2; \alpha, p_0 p_2) & \hbox{if} & x_1 < x_2 \\
S_{DW}(x_1; \alpha, p_0 p_1)S_{DW}(x_2; \alpha, p_2) & \hbox{if} & x_2 < x_1 \\
S_{DW}(x; \alpha, p_0 p_1 p_2) & \hbox{if} & x_1=x_2=x.
\end{array}%
\right.
\end{eqnarray}
The corresponding joint PMF of $(X_1,X_2)$ for $x_1 \in \mathbb{N}_o$ and $x_2 \in \mathbb{N}_0$ is given by
\begin{eqnarray}\label{r8}\nonumber
f_{X_1,X_2}(x_1,x_2)=\left\{%
\begin{array}{lll}
f_1(x_1,x_2) & \hbox{if} & 0\leq x_1<x_2 \\
f_2(x_1,x_2) & \hbox{if} & 0\leq x_2<x_1  \\
f_0 (x)& \hbox{if} & 0\leq x_1=x_2=x,\\
\end{array}%
\right.
\end{eqnarray}
where
\beanno
f_1(x_1,x_2) & = & f_{DW}(x_1; \alpha, p_1)f_{DW}(x_2; \alpha, p_0 p_2),  \\
f_2(x_1,x_2) & = & f_{DW}(x_1; \alpha, p_0 p_1)f_{DW}(x_2; \alpha, p_2),  \\
f_0(x) & = & u_1 f_{DW}(x; \alpha, p_0 p_2)- u_2 f_{DW}(x; \alpha, p_2),
\eeanno
in which $\ds u_1=p_1^{x^{\alpha}}$ and $\ds u_2=(p_0 p_1)^{{(x+1)}^{\alpha}}$.

The expressions $f_1(x_1, x_2)$, $f_2(x_1, x_2)$ and $f_0(x)$ for $x_1 \in \mathbb{N}_o$, $x_2 \in \mathbb{N}_0$ and $x \in \mathbb{N}_0$ have been obtained by means of the relation
$$
f_{X_1,X_2}(x_1,x_2) =
S_{X_1,X_2}(x_1,x_2) -S_{X_1,X_2}(x_1,x_2+1) - S_{X_1,X_2}(x_1+1,x_2) +S_{X_1,X_2}(x_1+1,x_2+1).
$$
The joint CDF of $(X_1, X_2)$ can be easily obtained from the following relation
$$
F_{X_1, X_2}(x_1, x_2) = F_{X_1}(x_1) + F_{X_2}(x_2) + S_{X_1, X_2}(x_1, x_2) - 1.
$$

\begin{figure}[h]
\begin{center}
\includegraphics[height=6cm,width=6cm]{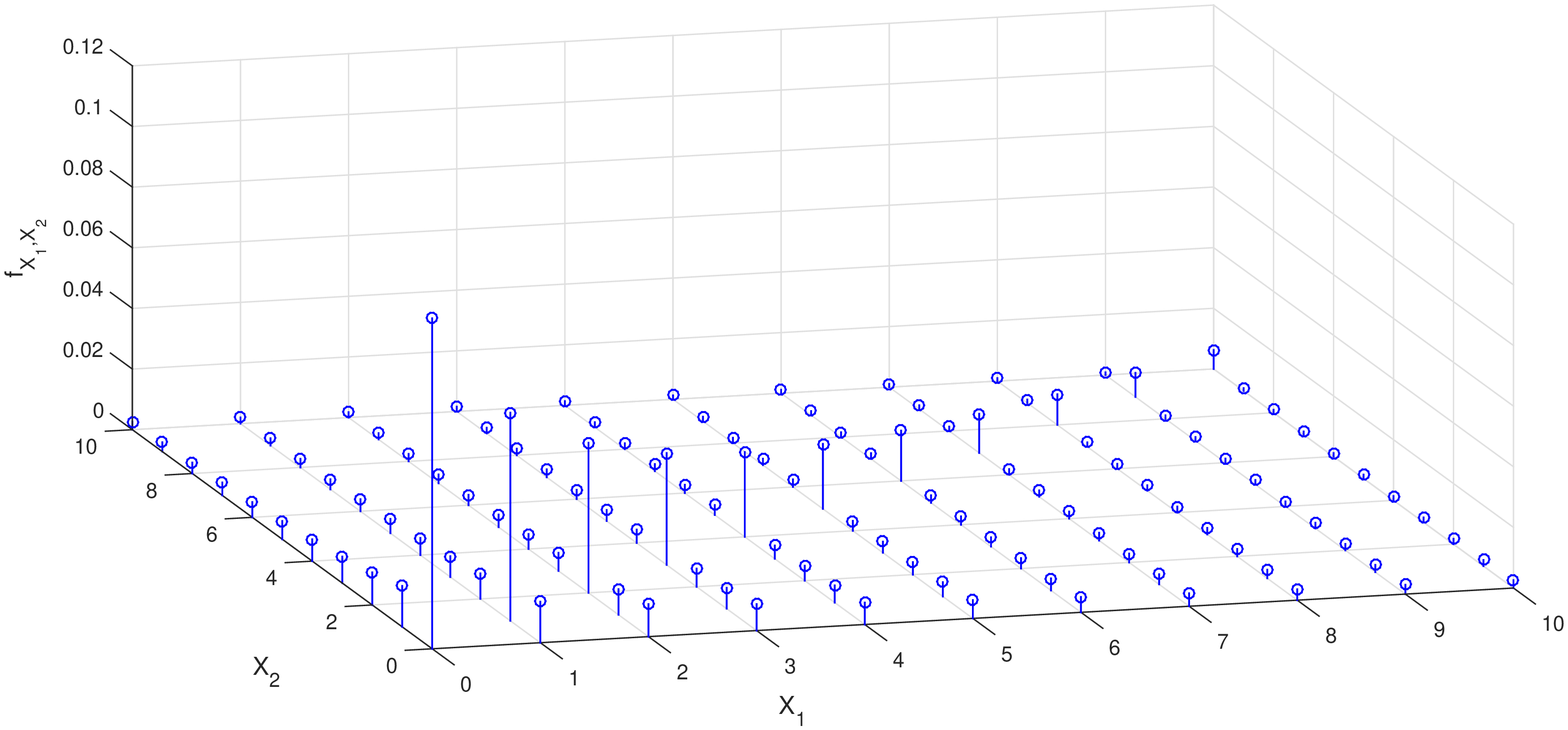}
\caption{The joint PMF of a BDW distribution when $\alpha = p_0 = p_1 = p_2$ = 0.9.  \label{pmf-1}}
\includegraphics[height=6cm,width=6cm]{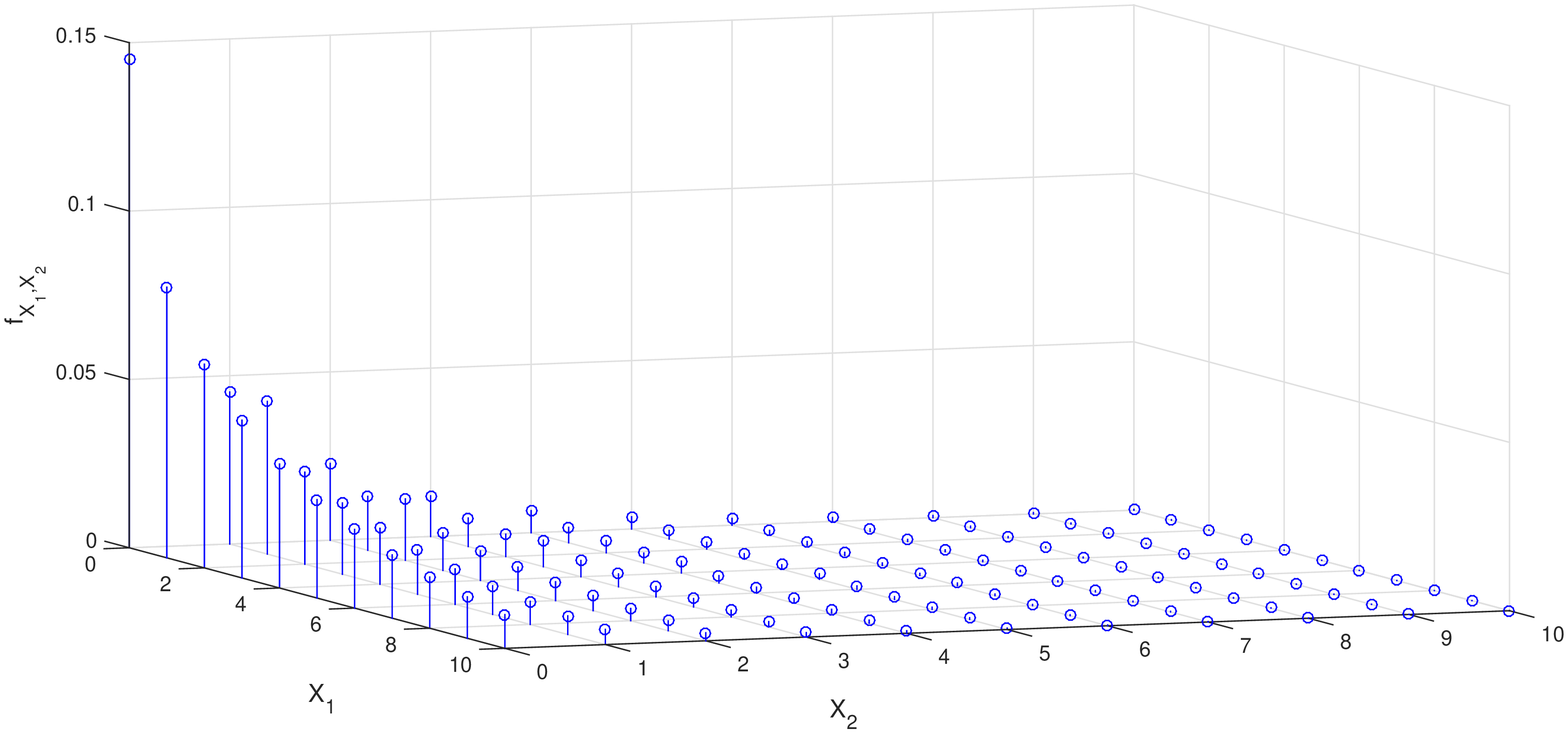}
\caption{The joint PMF of a BDW distribution when $\alpha$ = 0.9, $p_0$ = 0.95, $p_1$ = 0.8 and $p_2$ = 0.5.   \label{pmf-2}}
\end{center}
\end{figure}

In Figures \ref{pmf-1} and \ref{pmf-2} we have provided the plots of the joint PMF of BDW distributions for different parameter 
values.

\noindent The following interpretations can be provided for the BDW model.

\noindent {\sc Shock Model:} Suppose a system has two components, say Component 1 and Component 2.  It is assumed that
the system received shocks from three different sources, say Source A, Source B and Source C.  Each shock appears
randomly at discrete times, and independently of the other shocks.  Component 1 receives shocks from Source A and Source C, similarly,
Component 2 receives shocks from Source B and Source C.  A component fails as soon as it receives the first shock.  If
$U_A, U_B$ and $U_C$ denote the discrete times at which shocks appear from Source A, Source B and Source C, respectively, then
$X_1 = \min\{U_A, U_C\}$ and $X_2 = \min\{U_B, U_C\}$ denote the discrete lifetime of Component 1 and Component 2, respectively.
Therefore, if $U_A \sim$ DW$(\alpha,p_1)$, $U_B \sim$ DW$(\alpha, p_2)$ and $U_C \sim$ DW$(\alpha, p_0)$, then $(X_1, X_2) \sim$
BDW$(\alpha, p_0, p_1, p_2)$.

\noindent {\sc Masked Competing Risks Model:}  Suppose a system has two components, say Component 1 and Component 2.  Each component
can fail due to more than one causes.  Component 1 can fail due to Cause A and Cause C, similarly, Component 2
can fail due to Cause B and Cause C.  It is assumed that the failure times of the components, say $X_1$ and $X_2$ for Component 1 and
Component 2, respectively, are measured in discrete units and the causes of failures are masked.  Based on the Cox's latent failure time
model assumptions, see Cox \cite{Cox:1959}, if $U_1$, $U_2$ and $U_0$ denote lifetimes (in discrete units) due to Cause A, Cause B and
Cause C, respectively, then $X_1 = \min\{U_1, U_0\}$ and $X_2 = \min\{U_2, U_0\}$.  Therefore, in this case, if
$U_1 \sim$ DW$(\alpha,p_1)$, $U_2 \sim$ DW$(\alpha, p_2)$ and $U_0 \sim$ DW$(\alpha, p_0)$, then $(X_1, X_2) \sim$
BDW$(\alpha, p_0, p_1, p_2)$.

\subsection{\sc Properties}

First, note that if $(X_1,X_2)\sim$ BDW$(\alpha, p_0, p_1, p_2)$, then the marginals are DW distributions.  More precisely,
$X_1\sim$ DW$(\alpha, p_0 p_1)$ and $X_2\sim$ DW$(\alpha, p_0 p_2)$.  Moreover, it easily follows that if $(Y_1, Y_2) \sim$ MOBW$(\alpha,
\lambda_0, \lambda_1, \lambda_2)$, then $(X_1, X_2) \sim$ BDW$(\alpha, p_0, p_1, p_2)$, where $X_1 = [Y_1], X_2 = [Y_2]$ and
$p_0 = e^{-\lambda_0}$, $p_1 = e^{-\lambda_1}$, $p_2 = e^{-\lambda_2}$.  Therefore, the proposed BDW distribution can be considered as a natural discrete analogues of the continuous MOBW distribution.

We have also the following results regarding the conditional distributions of $X_1$ given $X_2$,
when $(X_1,X_2)\sim$ BDW$(\alpha, p_0, p_1, p_2)$.  The proofs are quite standard and the details are avoided.

\noindent {Proposition 2:} (a) The conditional PMF of $X_1$ given $X_2=x_2$, say $f_{X_1|X_2=x_2}(x_1|x_2)$, is given by
\begin{eqnarray}\label{r80}\nonumber
f_{X_1|X_2=x_2}(x_1|x_2)=\left\{%
\begin{array}{lll}
f_1(x_1|x_2) & \hbox{if} & 0\leq x_1<x_2  \\
f_2(x_1|x_2) & \hbox{if} & 0\leq x_2<x_1  \\
f_0 (x_1|x)& \hbox{if} & 0\leq x_1=x_2=x,
\end{array}%
\right.
\end{eqnarray}
where
$$
f_i(x_1|x_2)=\frac{f_i(x_1,x_2)}{f_{DW}(x_2; \alpha,p_0 p_2)}, \ \ \ i = 1, 2
$$
and
\begin{eqnarray}\nonumber
f_0(x_1|x)=\frac{f_0(x)}{f_{DW}(x;\alpha, p_0 p_2)} = p_1^{x^{\alpha}}-(p_0 p_1)^{(x+1)^{\alpha}}\frac{f_{DW}(x;\alpha,p_2)}
{f_{DW}(x;\alpha, p_0 p_2)}.
\end{eqnarray}
\noindent (b) The conditional SF of $X_1$ given $X_2\geq x_2$, say $S_{X_1|X_2\geq x_2}(x_1)$, is given by
\begin{eqnarray}\label{r9}\nonumber
S_{X_1|X_2\geq x_2}(x_1)&=&P(X_1\geq x_1|X_2\geq x_2)\\\nonumber
&=&\left\{%
\begin{array}{lll}
S_{DW}(x_1;\alpha, p_1) & \hbox{if} & 0\leq x_1<x_2  \\
S_{DW}(x_1; \alpha, p_0 p_1)/S_{DW}(x_2; \alpha,p_2) & \hbox{if} & 0\leq x_2<x_1   \\
S_{DW}(x;\alpha,p_1) & \hbox{if} & 0\leq x_1=x_2=x.
\end{array}%
\right.
\end{eqnarray}
(c) The conditional SF of $X_1$ given $X_2=x_2$, say $S_{X_1|X_2=x_2}(x_1)$, is given by
\begin{eqnarray}\label{r10}\nonumber
S_{X_1|X_2=x_2}(x_1)&=&P(X_1\geq x_1|X_2=x_2)\\\nonumber
&=&\left\{%
\begin{array}{lll}
S_{DW}(x_1;\alpha,p_1) & \hbox{if} & 0\leq x_1<x_2   \\
\frac{S_{DW}(x_1; \alpha, p_0 p_1)f_{DW}(x_2; \alpha,p_2)}{f_{DW}(x_2;\alpha,p_0 p_2)} & \hbox{if} & 0\leq x_2<x_1  \\
S_{DW}(x;\alpha,p_1) & \hbox{if} & 0\leq x_1=x_2=x.
\end{array}%
\right.
\end{eqnarray}

Now we show that if $(X_1,X_2)\sim$ BDW$(\alpha_1,\alpha_2,\alpha_3,p)$, then $X_1$ and $X_2$ are positive quadrant dependent.
First note that
$$
S_{X_1}(x_1)S_{X_2}(x_2) = (p_0 p_1)^{{x_1}^{\alpha}} (p_0 p_2)^{{x_2}^{\alpha}}.
$$
Hence, from (\ref{r6}) we obtain
$$
S_{X_1,X_2}(x_1,x_2)\geq S_{X_1}(x_1)S_{X_2}(x_2).
$$
In view of the fact that
$$
S_{X_1,X_2}(x_1,x_2)-S_{X_1}(x_1)S_{X_2}(x_2)=F_{X_1,X_2}(x_1,x_2)-F_{X_1}(x_1)F_{X_2}(x_2),
$$
it follows that for all values of $x_1\geq 0$ and $x_2\geq 0$,
$$
F_{X_1,X_2}(x_1,x_2)\geq F_{X_1}(x_1)F_{X_2}(x_2).
$$
Therefore, $X_1$ and $X_2$ are positive quadrant dependent. That is for every pair of increasing functions $m_1(.)$ and $m_2(.)$, it follows that Cov$(m_1(X_1),m_2(X_2)) \geq 0$; see for example Nelsen \cite{Nelsen:2006}.

Further observe that $X_1$ and $X_2$ are independent when $p_0$ = 0.  Therefore, in this case, Corr$(X_1, X_2) = 0$, for
fixed $\alpha$, $p_1$ and $p_2$.  Moreover, as $p_1 \rightarrow 1$ and $p_2 \rightarrow 1$, then $\lim_{p_1,p_2 \rightarrow 1}$
Corr$(X_1, X_2)$ = 1.  Hence, in a BDW distribution the correlation coefficient has the range $[0, 1)$.  In addition,
if $\alpha=1$, then $(X_1,X_2)$ has geometric marginals.  On the other hand, we have a new three-parameter bivariate geometric distribution with
parameters $p_0, p_1$  and $p_2$, whose joint SF is
\be
S_{X_1,X_2}(x_1,x_2) = p_1^{x_1}p_2^{x_2}p_0^{z}.
\ee
Here $x_1\in \mathbb{N}_0$, $x_2 \in \mathbb{N}_0$ and $z = \max\{x_1, x_2\}$ as before.  Moreover in this case $X_1$ and $X_2$
both have geometric distributions with parameter $p_0 p_1$ and $p_0 p_2$, respectively.  It may be mentioned that, recently,
Nekoukhou and Kundu \cite{NK:2017} obtained a two-parameter bivariate geometric distribution with joint CDF as
$$
F_{X_1,X_2}(x_1,x_2)=(1-p^{x_1+1})^{1-\alpha}(1-p^{x_2+1})^{1-\alpha}(1-p^{z+1})^{\alpha},
$$
where $0<p<1$, $\alpha>0$ and $z=\min\{x_1,x_2\}$.  

We have the following two results.

\noindent {Proposition 3:} Suppose $(X_1,X_2) \sim$ BDW$(\alpha, p_0, p_1, p_2)$, then $\min\{X_1, X_2\} \sim$
DW$(\alpha, p_0 p_1 p_2)$.

\noindent {\sc Proof:} The proof can be easily obtained by using the fact that
$$
P(\min\{X_1,X_2\}\geq x)=P(U_1\geq x, U_2\geq x, U_3\geq x)=(p_0 p_1 p_2)^{x^{\alpha}}.
$$

\noindent {Proposition 4:} Suppose $(X_{i1}, X_{i2}) \sim$ BDW$(\alpha, p_{i0}, p_{i1}, p_{i2})$, for $i = 1, \ldots, n$,
and they are independently distributed.  If $\ds Y_1 = \min\{X_{11}, \ldots, X_{n1}\}$ and $\ds Y_2 = \min\{X_{12}, \ldots, X_{n2}\}$,
then $(Y_1, Y_2) \sim$ BDW$\ds \left (\alpha, \prod_{i=1}^n p_{i0}, \prod_{i=1}^n p_{i1}, \prod_{i=1}^n p_{i2} \right )$.

\noindent {\sc Proof:} The proof can be easily obtained from the joint SF and, hence, the details are avoided. \qed

The joint probability generating function (PGF) of $X_1$ and $X_2$, for $|z_1| < 1$ and $|z_2| < 1$, can be
written as infinite mixtures,
\beanno
G_{X_1,X_2}(z_1, z_2)= E(z_1^{X_1} z_2^{X_2}) &=& \sum_{j=0}^{\infty} \sum_{i=0}^{\infty} P(X_1 = i, X_2 = j) z_1^i z_2^j  \\
& = & \sum_{j=0}^{\infty} \sum_{i=0}^{j-1} \left\{p^{i^{\alpha_1}}-p^{(i+1)^{\alpha_1}}\right\} \left\{p^{j^{\alpha_2+\alpha_3}}-p^{(j+1)^{\alpha_2+\alpha_3}}\right\}z_1^i z_2^j\\
& + &  \sum_{j=0}^{\infty} \sum_{i=j+1}^{\infty}
\left\{p^{i^{\alpha_1+\alpha_3}}-p^{(i+1)^{\alpha_1+\alpha_3}}\right\}\left\{p^{j^{\alpha_2}}-p^{(j+1)^{\alpha_2}}\right\}z_1^i z_2^j \\
& + & \sum_{i=0}^{\infty} p^{i^{\alpha_1}}\left\{p^{i^{\alpha_2+\alpha_3}}-p^{(i+1)^{\alpha_2+\alpha_3}}\right\} z_1^i z_2^i  \\
& - & \sum_{i=0}^{\infty} p^{(i+1)^{\alpha_1+\alpha_3}}\left\{p^{i^{\alpha_2}}-p^{(i+1)^{\alpha_2}}\right\} z_1^i z_2^i.
\eeanno
Hence, different moments and product moments of a BDW distribution can be obtained, as infinite series,
using the joint PGF.

Let us recall that a function $g(x,y): \mathbb{R} \times \mathbb{R} \rightarrow \mathbb{R}$,
is said to have a total positivity of order two (TP$_2$) property if $g(x,y)$ satisfies
\be
g(x_1,y_1) g(x_2, y_2) \ge g(x_2, y_1) g(x_1,y_2) \ \ \ \ \hbox{for all} \ \ \ \ x_1, y_1, x_2, y_2 \in \mathbb{R}.  \label{tp2}
\ee
\noindent {Proposition 5:} If $(X_1,X_2)\sim$ BDW$(\alpha, p_0, p_1, p_2)$, then the joint SF of $(X_1, X_2)$ satisfies
the TP$_2$ property.

\noindent {\sc Proof:} Suppose $x_{11}, x_{21}, x_{12}, x_{22} \in \mathbb{N}_0$ and $x_{11} < x_{21} < x_{12} < x_{22}$, then observe that
$$
\frac{S_{X_1,X_2}(x_{11},x_{21}) S_{X_1,X_2}(x_{12}, x_{22})}{S_{X_1,X_2}(x_{12},x_{21}) S_{X_1,X_2}(x_{11}, x_{22})} = p_0^{x_{21}^{\alpha} - x_{12}^{\alpha}} \ge 1.
$$
Similarly considering all other cases such as $x_{11} = x_{21} < x_{12} < x_{22}$, $x_{21} < x_{11} < x_{12} < x_{22}$ etc. it can be shown that
it satisfies (\ref{tp2}).  Hence, the result is proved.   \qed

It may be mentioned that TP$_2$ property is a very strong property and it ensures several ordering properties of the 
corresponding lifetime distributions, see for example Hu et al. \cite{HKS:2003} in this respect.  
Hence, the proposed BDW distribution satisfies those properties.

\section{\sc Maximum Likelihood Estimation}

In this section we consider the method of computing the MLEs of the unknown parameters based on a random
sample from BDW$(\alpha, p_0, p_1, p_2)$.  Suppose we have a random sample of size $n$ from a BDW$(\alpha, p_0, p_1, p_2)$ distribution as 
\be
{\cal D} = \{(x_{11}, x_{21}), \ldots, (x_{1n}, x_{2n})\}.   \label{data}
\ee
We use the following notations $I_1 = \{i: x_{1i} < x_{2i}\}$, $I_2 = \{i: x_{1i} > x_{2i}\}$ and $I_0 = \{i: x_{1i} = x_{2i} = x_i\}$,
and $n_j$ denotes the number of elements in the set $I_j$, for $j$ = 0, 1 and 2.  Now based on the observations (\ref{data}),
the log-likelihood function becomes
\bea
l(\alpha, p_0, p_1, p_2|{\cal D}) & = & \sum_{i \in I_1} \ln \left [ p_1^{x_{1i}^{\alpha}} - p_1^{(x_{1i}+1)^{\alpha}} \right ]  +
\sum_{i \in I_1} \ln \left [ (p_0 p_2)^{x_{2i}^{\alpha}} - (p_0 p_2)^{(x_{2i}+1)^{\alpha}} \right ] +  \nonumber \\
&  & \sum_{i \in I_2} \ln \left [ (p_0 p_1)^{x_{1i}^{\alpha}} - (p_0 p_1)^{(x_{1i}+1)^{\alpha}} \right ]  +
\sum_{i \in I_2} \ln \left [ p_2^{x_{2i}^{\alpha}} - p_2^{(x_{2i}+1)^{\alpha}} \right ]  +  \nonumber \\
&  & \sum_{i \in I_0} \ln \left [ p_1^{x_i^{\alpha}} \left ( (p_0 p_2)^{x_i^{\alpha}} - (p_0 p_2)^{(x_i+1)^{\alpha}} \right ) -
(p_0 p_1)^{(x_i+1)^{\alpha}} \left ( p_2^{x_i^{\alpha}} - p_2^{(x_i+1)^{\alpha}} \right ) \right ].    \nonumber  \\
\label{ll}
\eea
Hence, the MLEs of the unknown parameters can be obtained by maximizing (\ref{ll}) with respect to the unknown parameters.   It
involves solving a four dimensional optimization problem.  Clearly
analytical solutions do not exist.  Standard numerical methods like Newton-Raphson may be used to solve the optimization problem, but
it needs very good initial guesses.  Moreover, it is well known that it may converge to a local maximum rather than a global maximum.

To avoid that problems we propose to use EM algorithm to compute the MLEs in this case.  We mainly discuss about estimating
$\alpha$, $\lambda_0$, $\lambda_1$ and $\lambda_2$.  Kundu and
Dey \cite{KD:2009} developed a very efficient EM algorithm to compute the MLEs of the unknown parameters of a MOBW model.  At each
`E'-step the corresponding `M'-step can be performed by solving one non-linear equation only.  Kundu and Dey \cite{KD:2009}, by extensive simulation
experiments, indicated that the proposed EM algorithm converges to the global optimum solution and works very well even for moderate sample sizes.  Moreover,
if the shape parameter is known, then at the `M'-step the optimal solution can be obtained analytically.

In case of BDW model we have proposed the following EM algorithm, and because of its nested nature we call it as the nested EM
algorithm.  We treat this problem as a missing value problem.  It is assumed
that the complete data is of the form
$$
{\cal D}_c = \{(y_{11}, y_{21}), \ldots, (y_{1n}, y_{2n})\},
$$
where $\{(y_{1i}, y_{2i}); i = 1, \ldots, n\}$ is a random sample of size $n$ from MOBW$(\alpha, \lambda_0, \lambda_1, \lambda_2)$, and
$x_{1i} = [y_{1i}]$, $x_{2i} = [y_{2i}]$, for $i = 1, \ldots, n$.  We observe
$(x_{1i}, x_{2i})$ and $(y_{1i}, y_{2i})$ is missing.  At each step we estimate the missing values by maximized likelihood principle
method.  The following result will be useful for that purpose.

\noindent {\sc Theorem 1:} Suppose $(Y_1, Y_2) \sim$ MOBW$(\alpha, \lambda_0, \lambda_1 \lambda_2)$, $Y = \min\{Y_1, Y_2\}$,
and $X_1 = [Y_1]$, $X_2 = [Y_2]$. Then, the conditional PDF of $(Y_1, Y_2)$ given $(X_1, X_2)$ is

\noindent (a) If $i < j$, and $i \le y_1 < i+1, j \le y_2 < j+1$, then
$$
f_{Y_1, Y_2}(y_1, y_2|X_1 = i, X_2 = j) =
\frac{f_{WE}(y_1; \alpha, \lambda_1) f_{WE}(y_2; \alpha, \lambda_0+\lambda_2)}
{P(i \le Y_1 < i+1, j \le Y_2 < j+1)}
$$
and zero, otherwise.

\noindent (b) If $i > j$, and $i \le y_1 < i+1, j \le y_2 < j+1$, then
$$
f_{Y_1, Y_2}(y_1, y_2|X_1 = i, X_2 = j) =
\frac{f_{WE}(y_1; \alpha, \lambda_0 + \lambda_1) f_{WE}(y_2; \alpha, \lambda_2)}
{P(i \le Y_1 < i+1, j \le Y_2 < j+1)}
$$
and zero, otherwise.

\noindent (c) If $i = j$, and $i \le y_1 = y_2 = y < i+1$, then
$$
f_{Y_1, Y_2}(y|X_1 = i, X_2 = i) =
\frac{f_{WE}(y_1; \alpha, \lambda_0+\lambda_1+\lambda_2)}
{P(i \le Y < i+1)}
$$
and zero, otherwise.

\noindent (d) If $i = j$, and $i \le y_1 < y_2 < i+1$, then
$$
f_{Y_1, Y_2}(y_1, y_2|X_1 = i, X_2 = i) =
\frac{f_{WE}(y_1; \alpha, \lambda_1) f_{WE}(y_2; \alpha, \lambda_0+\lambda_2)}
{P(i \le Y_1 < i+1, i \le Y_2 < i+1)}
$$
and zero, otherwise.

\noindent (e) If $i = j$, and $i \le y_2 <  y_1 < i+1$, then
$$
f_{Y_1, Y_2}(y_1, y_2|X_1 = i, X_2 = i) =
\frac{f_{WE}(y_1; \alpha, \lambda_0 + \lambda_1) f_{WE}(y_2; \alpha, \lambda_2)}
{P(i \le Y_1 < i+1, i \le Y_2 < i+1)}
$$
and zero, otherwise.

\noindent {\sc Proof:} The proof can be easily obtained by using conditioning argument, and the details are avoided.    \qed

Based on Theorem 1, if $(Y_1, Y_2) \sim$ MOBW$(\alpha, \lambda_0, \lambda_1 \lambda_2)$, and
$X_1 = [Y_1]$, $X_2 = [Y_2]$, then for known $\alpha$, $\lambda_0$, $\lambda_1$ and $\lambda_2$, the maximum likelihood predictor of
$(Y_1, Y_2)$ given $X_1 = i$ and $X_2 = j$, say $(\widehat{Y}_1, \widehat{Y}_2)$, can be easily obtained.  The explicit expressions of
$\widehat{Y}_1$ and $\widehat{Y}_2$ are provided in the Appendix.  Note that $\widehat{Y}_1$ and $\widehat{Y}_2$ depend on
$\alpha, \lambda_0, \lambda_1 \lambda_2$, and $i,j$, but we are not making it explicit.

Now we propose the following nested EM algorithm to compute the MLEs of the unknown parameters.

\noindent {\sc Algorithm 1: Nested EM Algorithm}

\begin{itemize}
\item Suppose at the $k$-th step of the outer EM algorithm the estimates $\alpha$, $\lambda_0$, $\lambda_1$ and $\lambda_2$, are
$\alpha^{(k)}$, $\lambda_0^{(k)}$, $\lambda_1^{(k)}$ and $\lambda_2^{(k)}$, respectively.

\item For the given $\alpha^{(k)}$, $\lambda_0^{(k)}$, $\lambda_1^{(k)}$ and $\lambda_2^{(k)}$, based on maximized
likelihood principle as discussed above obtain ${\cal D}_c^{(k)} =
\{(\widehat{y}_{11}, \widehat{y}_{21}), \ldots, (\widehat{y}_{11}, \widehat{y}_{21})$ from ${\cal D}$.

\item Based on ${\cal D}_c^{(k)}$, using the EM algorithm proposed by Kundu and Dey \cite{KD:2009}, obtain
$\alpha^{(k+1)}$, $\lambda_0^{(k+1)}$, $\lambda_1^{(k+1)}$ and $\lambda_2^{(k+1)}$.

\item Continue the process until the convergence takes place.

\end{itemize}

Once the MLEs of the unknown parameters are obtained, then at the last stage of the outer EM, using the method of Louis \cite{Louis:1982}
the confidence intervals of the unknown parameters can be obtained.  One of the natural questions is how to obtain the initial estimates
of the unknown parameters.  Since $X_1 \sim$ DW$(\alpha, p_0 p_1)$, $X_2 \sim$ DW$(\alpha, p_0 p_2)$ and $\min\{X_1, X_2\} \sim$
DW$(\alpha, p_0 p_1 p_2)$, from $\{x_{1i}; i = 1, \ldots, n\}$, $\{x_{2i}; i = 1, \ldots, n\}$ and $\{\min\{x_{1i},x_{2i}\};
i = 1, \ldots, n\}$, we can obtain initial estimates of $\alpha$, $p_0$, $p_1$ and $p_2$.  The details will be explained in the Data
Analysis section.

\section{\sc Bayes Estimation}

In this section we obtain the Bayes estimates of $\alpha$, $\lambda_0$, $\lambda_1$ and $\lambda_2$ based on a random sample of size
$n$ as described in (\ref{data}).  It is assumed that $\lambda_0$, $\lambda_1$ and $\lambda_2$ has a Dirichlet-Gamma prior as described
in (\ref{prior}).  We do not assume any specific form of prior on $\alpha$.  It is simply assumed that the support of $\alpha$ is
$(0, \infty)$, and it has the PDF which is log-concave.  Moreover, the prior on $\alpha$ and $\lambda_0, \lambda_1, \lambda_2$ are
independently distributed.  Let us denote ${\ve \theta} = (\alpha, \lambda_0, \lambda_1, \lambda_2)$, and the joint prior on ${\ve \theta}$
as $\pi(\ve \theta)$.  In view of the fact that the discrete case is considered, the posterior distribution of ${\ve \theta}$, say
$\pi({\ve \theta}|{\cal D})$, is not so easy to handle computationally.  In a situation like this, Ghosh et al. \cite{GSD:2006}
(Chapter 7) suggested to use some data augmentation method  which might help.

Recently Kundu and Gupta \cite{KG:2013} provided a very efficient method to compute the Bayes estimates
and the associated highest posterior density (HPD) credible intervals of $\alpha$, $\lambda_0$, $\lambda_1$ and $\lambda_2$ with respect
to the above priors and based on a
random sample of size $n$ from MOBW$(\alpha, \lambda_0, \lambda_1, \lambda_2)$.  If the shape parameter $\alpha$ is known, then
Dirichlet-Gamma prior becomes a conjugate prior and in this case the Bayes estimates and the associated credible intervals of
$\lambda_0$, $\lambda_1$ and $\lambda_2$ can be obtained in explicit forms.  If the shape parameter is unknown, then a very efficient
Gibbs sampling technique has been proposed by Kundu and Gupta \cite{KG:2013} and that can be used to compute the Bayes estimates
and the associated HPD credible intervals.  In case of BDW distribution to compute the Bayes estimates of the unknown parameters, we
have combined the `data augmentation' method as suggested by Ghosh et al. \cite{GSD:2006} and the efficient Gibbs sampling method as
suggested by Kundu and Gupta \cite{KG:2013} in case MOBW distribution.  We propose the following algorithm to compute the Bayes estimates
and the associated HPD credible intervals of any function of $\alpha$, $\lambda_0$, $\lambda_1$ and $\lambda_2$, say
$g(\alpha, \lambda_0, \lambda_1, \lambda_2)$, based on the random sample (\ref{data}).

\noindent {\sc Algorithm 2: Augmented-Gibbs Sampling Procedure}

\noindent Step 1: Obtain initial estimates of $\alpha$, $\lambda_0$, $\lambda_1$ and $\lambda_2$, say ${\ve \theta}^{(0)} = (\alpha^{(0)}, \lambda_0^{(0)},
\lambda_1^{(0)}, \lambda_2^{(0)})$.

\noindent Step 2: Based on ${\ve \theta}^{(0)}$ obtain ${\cal D}^{(0)} = \{(y_{11}^{(0)},y_{21}^{(0)}, \ldots, (y_{1n}^{(0)},y_{2n}^{(0)})\}$ as suggested in
the previous section by using maximized likelihood principle.

\noindent Step 3: Using the augmented data ${\cal D}^{(0)}$ and using the Gibbs sampling method suggested by Kundu and Gupta \cite{KG:2013} generate
$\{{\ve \theta}^{(0i)} = (\alpha^{(0i)}, \lambda_0^{(0i)}, \lambda_1^{(0i)}, \lambda_2^{(0i)}); i = 1, \ldots M\}$.

\noindent Step 4: Obtain ${\ve \theta}^{(1)} = (\alpha^{(1)}, \lambda_0^{(1)}, \lambda_1^{(1)}, \lambda_2^{(1)})$, where
$$
\alpha^{(1)} = \frac{1}{M}\sum_{i=1}^M \alpha^{(0i)}, \ \ \lambda_0^{(1)} = \frac{1}{M}\sum_{i=1}^M \lambda_0^{(0i)}, \ \
\ \ \lambda_1^{(1)} = \frac{1}{M}\sum_{i=1}^M \lambda_1^{(0i)}, \ \ \ \ \lambda_2^{(1)} = \frac{1}{M}\sum_{i=1}^M \lambda_2^{(0i)}.
$$

\noindent Step 5: Go back to Step 1 and replace ${\ve \theta}^{(0)}$ by ${\ve \theta}^{(1)}$ and continue the process $N$ times.

\noindent Step 6: At the $N$-th step we obtain the generated samples
\be
\{{\ve \theta}^{(Ni)} = (\alpha^{(Ni)}, \lambda_0^{(Ni)}, \lambda_1^{(Ni)}, \lambda_2^{(Ni)}); i = 1, \ldots M\}.   \label{gens}
\ee
Based on the generated samples (\ref{gens}) we can easily compute a simulation consistent Bayes estimate of $g(\alpha, \lambda_0,
\lambda_1, \lambda_2)$ as
$$
\widehat{g}_B(\alpha, \lambda_0, \lambda_1, \lambda_2) = \frac{1}{M} \sum_{i=1}^M g(\alpha^{(Ni)}, \lambda_0^{(Ni)}, \lambda_1^{(Ni)}, \lambda_2^{(Ni)}).
$$

\noindent Step 7: If we denote
$$
g_i = g(\alpha^{(Ni)}, \lambda_0^{(Ni)}, \lambda_1^{(Ni)}, \lambda_2^{(Ni)}), \ \ \ i = 1, \ldots, M,
$$
and $g_{(1)} < g_{(2)} < \ldots < g_{(N)}$ denote the ordered $g_i$'s, then based on $g_{(i)}$'s in a routine manner we can construct
100(1-$\beta$)\% credible and HPD credible intervals of $g(\alpha, \lambda_0, \lambda_1, \lambda_2)$, see for example Kundu and 
Gupta \cite{KG:2013}.

\section{\sc Data Analysis}

\subsection{\sc Football Data}

In this section we present the analysis of a data set to see how the proposed model and methods can be applied in practice.
The data set which we have analyzed here represents the Italian Series A football match score played between two Italian football giants
`ACF Firontina' ($X_1$) and `Juventus' ($X_2$) during the period 1996 to 2011.  The data set is presented below.
\begin{table}[h]
\bc
\begin{tabular}{|l|c|c|l||c|c|}  \cline{1-6}

\hline
Obs. & ACF  & Juventus & Obs.   & ACF & Juventus  \\
     & Firontina  &    &    &    Firontina    &           \\
     &   ($X_1$)  & ($X_2$) &  &  ($X_1$)  & ($X_2$)    \\
   &   &   &   &   &   \\   \hline \hline
1 & 1 & 2 & 14 & 1 & 2    \\
2 & 0 & 0 & 15 & 1 & 1    \\
3 & 1 & 1 & 16 & 1 & 3    \\
4 & 2 & 2 & 17 & 3 & 3    \\
5 & 1 & 1 & 18 & 0 & 1    \\
6 & 0 & 1 & 19 & 1 & 1    \\
7 & 1 & 1 & 20 & 1 & 2    \\
8 & 3 & 2 & 21 & 1 & 0    \\
9 & 1 & 1 & 22 & 3 & 0    \\
10 & 2 & 1 & 23 & 1 & 2   \\
11 & 1 & 2 & 24 & 1 & 1   \\
12 & 3 & 3 & 25 & 0 & 1   \\
13 & 0 & 1 & 26 & 0 & 1   \\    \hline
\end{tabular}
\ec
\caption{UEFA Champion's League data \label{football-data}}
\end{table}
First we have fitted DW distribution to $X_1$, $X_2$ and $\min\{X_1, X_2\}$.  The MLEs of $\alpha$ and $p$, and the results are presented in Table \ref{one-d-res}.

\begin{table}[h]
\bc
\begin{tabular}{|c|c|l|c|c|}  \cline{1-5}

\hline
Data & $\widehat{\alpha}$ & $\widehat{p}$ & $\chi^2$ & $p$-value   \\  \hline
$X_1$ & 1.8424 & 0.7617 & 5.5556 & 0.14    \\  \hline
$X_2$ & 2.4646 & 0.8604 & 0.8787 & 0.83   \\  \hline
min$\{X_1, X_2\}$ & 1.8398 & 0.6818 & 3.1301 & 0.37  \\  \hline
\end{tabular}
\ec
\caption{MLEs, chi-square and associated $p$-values for $X_1$, $X_2$ and min$\{X_1, X_2\}$.   \label{one-d-res}}
\end{table}

Based on the chi-square statistic and the associated $p$-values it seems that DW distribution fits $X_1$, $X_2$ and $\min\{X_1, X_2\}$
reasonably well.  We would like to fit BDW distribution to the above data set.  We have used the following initial estimates of the
unknown parameters,  
$$
\alpha^{(0)} = 2.0489, \ \ \lambda_0^{(0)} = 0.0395, \ \ \lambda_1^{(0)} = 0.2326, \ \
\lambda_2^{(0)} = 0.1108.
$$
From Table \ref{one-d-res} we obtain $\alpha^{(0)}$ by taking the average of the three estimates of $\alpha$ namely 1.8424, 
2.4646 and 1.8398, respectively.  Similarly, $\lambda_i^{(0)}$'s
are obtained by solving $p_i$'s uniquely from the three estimates of $p$, namely $p_0^{(0)} p_1^{(0)}$ = 0.7617, $p_0^{(0)} p_2^{(0)}$ = 
0.8604, $p_0^{(0)} p_1^{(0)} p_2^{(0)}$ = 0.6818, and using $p_i = e^{-\lambda_i}$, for $i$ = 0, 1 and 2.  

We start the EM algorithm with the above initial guesses.  We use the stopping criterion when the difference between the two consecutive 
pseudo log-likelihood values is less than $10^{-4}$.  The EM algorithm stops after 23 iterations and we obtain the MLEs and the associated 95\% confidence intervals
of the parameters as: $\ds \widehat{\alpha}_{MLE} = 4.9798 (\mp 0.8112)$, $\widehat{\lambda}_{0, MLE} = 0.0013 (\mp 0.0002)$, $\ds
\widehat{\lambda}_{1, MLE} = 0.2468 (\mp 0.0511)$ and $\ds \widehat{\lambda}_{2, MLE} = 0.0487 (0.0086)$.
To observe whether the proposed model provides a good fit to the data, we have obtained the chi-squared
statistic.  The observed $\chi^2$-value is 10.9690, with the $p$-value greater than 0.27, for the $\chi^2$ distribution with 9 degrees
of freedom.  Hence, it is clear that the proposed model and the nested EM algorithm work quite well in this case.  

Now for comparison 
purposes we want to see whether bivariate discrete exponential (BDE) fits the data or not.  Note that BDE can be obtained as a special 
case of the BDW when the common shape parameter is 1.  Hence, we want to perform the following test
$$
H_0: \alpha = 1 \ \ \ \ \hbox{vs.} \ \ \ \ H_1: \alpha \ne 1.
$$
Now based on the above 95\% confidence interval of $\alpha$, we can conclude that $H_0$ is rejected with 5\% level of significance.  
Hence, BDE cannot be used for this data set.

Now to compute the Bayes estimates and the associated HPD credible intervals we have used the following hyper-parameter of the
Dirichlet-Gamma prior: $a = b = a_0 = a_1 = a_2$ = 0.0001, and for $\pi_2(\alpha)$ it is assumed that it follows a gamma distribution
with the shape parameter $c$ = 0.0001 and the scale parameter $d$ = 0.0001.  The above hyper-parameters behave like non-informative priors
but they are still proper priors, see for example Congdon \cite{Congdon:2006}.  Based on the above hyper-parameters with 10,000 replications
we obtain the Bayes estimates and the associated 95\% HPD credible intervals as follows:
$\ds \widehat{\alpha}_{BE} = 4.3716 (\mp 0.7453)$, $\widehat{\lambda}_{0, BE} = 0.0019 (\mp 0.0002)$, $\ds
\widehat{\lambda}_{1, BE} = 0.2723 (\mp 0.0416)$ and $\ds \widehat{\lambda}_{2, BE} = 0.0318 (0.0093)$.  It is clear that the Bayes estimates
with respect to the non-informative priors and the MLEs behave very similarly.

\subsection{\sc Nasal Drainage Severity Score}

In this case the data represents the efficacy of steam inhalation in the treatment of common cold symptoms.  The patients had common
cold of recent onset.  Each patient has been given two 2-minutes steam inhalation treatment, after which severity of nasal drainage
was self assessed for the next four days.  The outcome variable at each day was ordinal with four categories:
0 = no symptoms; 1 = mild symptoms; 2 = moderate symptoms; 3 = severe symptoms.  We analyze the data for the first two days and they
are presented in Table \ref{nasal-data}.  The original data are available in Davis \cite{Davis:2002}.
\begin{table}[h]
\bc
\begin{tabular}{|l|c|c|l||c|c|}  \cline{1-6}

\hline
No. & Day 1 & Day 2 & No. & Day 1 & Day 2  \\
     &   &    &    &        &           \\
     &   ($X_1$)  & ($X_2$) &  &  ($X_1$)  & ($X_2$)    \\
   &   &   &   &   &   \\   \hline \hline
1 & 1 & 1 & 16 & 2 & 1    \\
2 & 0 & 0 & 17 & 1 & 1    \\
3 & 1 & 1 & 18 & 2 & 2    \\
4 & 1 & 1 & 19 & 3 & 1    \\
5 & 0 & 2 & 20 & 1 & 1    \\
6 & 2 & 0 & 21 & 2 & 1    \\
7 & 2 & 2 & 22 & 2 & 2    \\
8 & 1 & 1 & 23 & 1 & 1    \\
9 & 3 & 2 & 24 & 2 & 2    \\
10 & 2 & 2 & 25 & 2 & 0   \\
11 & 1 & 0 & 26 & 1 & 1   \\
12 & 2 & 3 & 27 & 0 & 1   \\
13 & 1 & 3 & 28 & 1 & 1   \\
14 & 2 & 1 & 29 & 1  & 1    \\
15 & 2 & 3 & 30 & 3 & 3 \\  \hline
\end{tabular}
\ec
\caption{Nasal drainage severity score for 30 patients.   \label{nasal-data}}
\end{table}

In this case, we have also fitted the DW distribution to $X_1$, $X_2$ and $\min\{X_1, X_2\}$, and the results are presented
in Table \ref{one-d-res-2}.  From the $p$-values in Table \ref{one-d-res-2} it is clear that DW fits $X_1$, $X_2$ and
$\min\{X_1, X_2\}$ very well.  Hence, it is reasonable to fit BDW to this data set.

\begin{table}[h]
\bc
\begin{tabular}{|c|c|l|c|c|}  \cline{1-5}

\hline
Data & $\widehat{\alpha}$ & $\widehat{p}$ & $\chi^2$ & $p$-value   \\  \hline
$X_1$ & 2.8280 & 0.9057 & 0.0366 & 0.99    \\  \hline
$X_2$ & 2.2768 & 0.8419 & 1.5676 & 0.67   \\  \hline
min$\{X_1, X_2\}$ & 2.4717 & 0.8031 & 0.0124 & 0.99  \\  \hline
\end{tabular}
\ec
\caption{MLEs, chi-square and associated $p$-values for $X_1$, $X_2$ and min$\{X_1, X_2\}$.   \label{one-d-res-2}}
\end{table}

We have used the proposed augmented-EM algorithm to compute the MLEs of the unknown parameters.  We have used the following initial
values to start the EM algorithm,
$$
\alpha^{(0)} = 2.5255, \ \ \lambda_0^{(0)} = 0.0519, \ \ \lambda_1^{(0)} = 0.0471, \ \
\lambda_2^{(0)} = 0.1202.
$$
We have used the same stopping criterion as before, and the EM algorithm stops after 15 iterations.  The MLEs and the associated
95\% confidence intervals are as follows: $\widehat{\alpha}_{MLE}$ = 3.6571 ($\mp$ 0.9787), $\widehat{\lambda}_{0,MLE}$ = 0.0699
($\mp$ 0.0178), $\widehat{\lambda}_{1, MLE}$ = 0.0025 ($\mp$ 0.0007), $\widehat{\lambda}_{2, MLE}$ = 0.0697 ($\mp$ 0.0156).  The
associated $\chi^2$ value becomes 13.6321 with the $p$-value greater than 0.13 for a $\chi^2$ distribution with 9 degrees of
freedom.  It clearly indicates that the proposed BDW distribution fits the bivariate nasal
drainage data set quite well.  Moreover, similarly as the previous data set, based on the confidence interval of $\alpha$ 
we can conclude that BDE cannot be used for this data set also.

In this case, we have also calculated the Bayes estimates using the same prior assumptions and
the same hyper-parameters as the previous example.  The Bayes estimates and the associated 95\% HPD credible intervals are
provided below:  $\widehat{\alpha}_{BE}$ = 3.7781 ($\mp$ 0.9321), $\widehat{\lambda}_{0,BE}$ = 0.0754
($\mp$ 0.0132), $\widehat{\lambda}_{1, BE}$ = 0.0017 ($\mp$ 0.0008), $\widehat{\lambda}_{2, BE}$ = 0.0721 ($\mp$ 0.0137).  In this
case, it is also observed that the MLEs and the Bayes estimates with respect to non-informative priors behave in a very similar
manner.

\section{\sc Conclusions}

In this paper we have introduced BDW distribution from three univariate DW distributions and using the minimization technique.
It is observed that the proposed BDW distribution has univariate DW marginals.  The proposed BDW distribution has four parameters
and due to which it becomes a very flexible bivariate discrete distribution. It has some interesting physical interpretations in terms of
shock model and latent failure time competing risks model.  It is observed the BDW distribution has the correlation range
$[0, 1)$ and it has the TP$_2$ property.  The MLEs cannot be obtained in explicit forms, and we have used nested EM algorithm
to compute the MLEs of the unknown parameters.    We have
also proposed augmented Gibbs sampling procedure to compute the Bayes estimates of the unknown parameters.  Two real data sets
have been analyzed for illustrative purposes.  It is observed that the nested EM algorithm and augmented Gibbs sampling method
work quite well in practice.

\section*{\sc Acknowledgements:} The authors would like to thank two unknown reviewers for their constructive comments which have 
helped us to improve the manuscript significantly.  The second author was partially supported by the grant Khansar-CMC-101.

\section*{\sc Appendix:}

In this Appendix, we provide the explicit expressions of $\widehat{Y}_1$ and $\widehat{Y}_2$. First, let us consider the function
$$
g(\alpha, \lambda) = \left ( \frac{\alpha-1}{\alpha \lambda} \right )^{1/\alpha},
$$
for $\alpha > 1$ and $\lambda > 0$;

\noindent (a) If $i < j$, then
\beanno
\widehat{Y}_1 & = & \left \{
\begin{array}{lll}
i & \hbox{if} & \alpha \le 1 \ \ \hbox{or} \ \ \ \alpha > 1 \ \ \hbox{and} \ \  g(\alpha, \lambda_1) < i   \\
g(\alpha, \lambda_1) & \hbox{if} & \alpha > 1 \ \ \hbox{and} \ \ i \le g(\alpha, \lambda_1) \le i+1 \\
i+1 & \hbox{if} & \alpha > 1 \ \ \hbox{and} \ \ g(\alpha, \lambda_1) > i+1,
\end{array}
\right .   \\
\widehat{Y}_2 & = & \left \{
\begin{array}{lll}
j & \hbox{if} & \alpha \le 1 \ \ \hbox{or} \ \ \ \alpha > 1 \ \ \hbox{and} \ \  g(\alpha, \lambda_0+\lambda_2) < j   \\
g(\alpha, \lambda_0+\lambda_2) & \hbox{if} & \alpha > 1 \ \ \hbox{and} \ \ i \le g(\alpha, \lambda_0+\lambda_2) \le j+1 \\
j+1 & \hbox{if} & \alpha > 1 \ \ \hbox{and} \ \ g(\alpha, \lambda_0+\lambda_2) > j+1.
\end{array}
\right .
\eeanno

\noindent (b) If $i > j$, then
\beanno
\widehat{Y}_1 & = & \left \{
\begin{array}{lll}
i & \hbox{if} & \alpha \le 1 \ \ \hbox{or} \ \ \ \alpha > 1 \ \ \hbox{and} \ \  g(\alpha, \lambda_0+\lambda_1) < i   \\
g(\alpha, \lambda_0+\lambda_1) & \hbox{if} & \alpha > 1 \ \ \hbox{and} \ \ i \le g(\alpha, \lambda_0+\lambda_1) \le i+1 \\
i+1 & \hbox{if} & \alpha > 1 \ \ \hbox{and} \ \ g(\alpha, \lambda_0+\lambda_1) > i+1,
\end{array}
\right .  \\
\widehat{Y}_2 & = & \left \{
\begin{array}{lll}
j & \hbox{if} & \alpha \le 1 \ \ \hbox{or} \ \ \ \alpha > 1 \ \ \hbox{and} \ \  g(\alpha, \lambda_2) < j   \\
g(\alpha, \lambda_2) & \hbox{if} & \alpha > 1 \ \ \hbox{and} \ \ i \le g(\alpha, \lambda_2) \le j+1 \\
j+1 & \hbox{if} & \alpha > 1 \ \ \hbox{and} \ \ g(\alpha, \lambda_2) > j+1.
\end{array}
\right .
\eeanno

\noindent (c) In the case $i = j$, in order to compute $\widehat{Y}_1$ and $\widehat{Y}_2$ we use the following notations,
$$
\widehat{W}   =  \left \{
\begin{array}{lll}
i & \hbox{if} & \alpha \le 1 \ \ \hbox{or} \ \ \ \alpha > 1 \ \ \hbox{and} \ \  g(\alpha, \lambda_0+\lambda_1+\lambda_2) < i   \\
g(\alpha, \lambda_0+\lambda_1+\lambda_2) & \hbox{if} & \alpha > 1 \ \ \hbox{and} \ \ i \le g(\alpha, \lambda_0+\lambda_1+\lambda_2)
\le i+1 \\
i+1 & \hbox{if} & \alpha > 1 \ \ \hbox{and} \ \ g(\alpha, \lambda_0+\lambda_1+\lambda_2) > i+1,
\end{array}
\right .
$$
and
$$
A  =  \frac{f_{WE}(\widehat{W}; \alpha, \lambda_0+\lambda_1+\lambda_2)}{P(i \le Y < i+1)}.
$$

If $\alpha \le 1$, then define $U_1 = U_2 = i$.  If $\alpha > 1$ and $g(\alpha, \lambda_1) < g(\alpha, \lambda_0+\lambda_2)$, then
define $U_1$ and $U_2$ as follows,
\beanno
\widehat{U}_1 & = & \left \{
\begin{array}{lll}
i & \hbox{if} & g(\alpha, \lambda_1) < i   \\
g(\alpha, \lambda_1) & \hbox{if} & i \le g(\alpha, \lambda_1) \le i+1 \\
i+1 & \hbox{if} & g(\alpha, \lambda_1) > i+1,
\end{array}
\right .  \\
\widehat{U}_2 & = & \left \{
\begin{array}{lll}
i & \hbox{if} & g(\alpha, \lambda_0+\lambda_2) < i   \\
g(\alpha, \lambda_0+\lambda_2) & \hbox{if} & i \le g(\alpha, \lambda_0+\lambda_2) \le i+1 \\
i+1 & \hbox{if} & g(\alpha, \lambda_0+\lambda_2) > i+1,
\end{array}
\right .
\eeanno
and
$$
B = \frac{f_{WE}(\widehat{U}_1; \alpha, \lambda_1) f_{WE}(\widehat{U}_2; \alpha, \lambda_0+\lambda_2)}{P(i \le Y_1 < i+1, i \le Y_2 < i+1)}.
$$

If $\alpha \le 1$, then define $V_1 = V_2 = i$.  If $\alpha > 1$ and $g(\alpha, \lambda_2) < g(\alpha, \lambda_0+\lambda_1)$, then
define $V_1$ and $V_2$ as follows,
\beanno
\widehat{V}_1 & = & \left \{
\begin{array}{lll}
i & \hbox{if} & g(\alpha, \lambda_0+\lambda_1) < i   \\
g(\alpha, \lambda_0+\lambda_1) & \hbox{if} & i \le g(\alpha, \lambda_0+\lambda_1) \le i+1 \\
i+1 & \hbox{if} & g(\alpha, \lambda_0+\lambda_1) > i+1,
\end{array}
\right .  \\
\widehat{V}_2 & = & \left \{
\begin{array}{lll}
i & \hbox{if} & g(\alpha, \lambda_2) < i   \\
g(\alpha, \lambda_2) & \hbox{if} & i \le g(\alpha, \lambda_2) \le i+1 \\
i+1 & \hbox{if} & g(\alpha, \lambda_2) > i+1,
\end{array}
\right .
\eeanno
and
$$
C = \frac{f_{WE}(\widehat{V}_1; \alpha, \lambda_0+\lambda_1) f_{WE}(\widehat{V}_2; \alpha, \lambda_2)}{P(i \le Y_1 < i+1, i \le Y_2 < i+1)}.
$$
Therefore, we have
$$
(\widehat{Y}_1, \widehat{Y}_2) = \left \{
\begin{array}{lll}
(\widehat{W}, \widehat{W}) & \hbox{if} & A > \max\{B, C\}  \\
(\widehat{U}_1, \widehat{U}_2) & \hbox{if} & B > \max\{A, C\}  \\
(\widehat{V}_1, \widehat{V}_2) & \hbox{if} & C > \max\{A, B\}.  \\
\end{array}
\right .
$$

\end{document}